# A Study on Internet of Things based Applications

Deeksha Jain, P. Venkata Krishna and V. Saritha

School of Computing Science and Engineering

VIT University, Vellore, TN, India

*Abstrac*t

*This paper gives a detail analysis of various applications based on Internet of Thing (IoT)s. This explains about how internet of things evolved from mobile computing and ubiquitous computing. It emphasises the fact that objects are connected over the internet rather than people. The properties of Internet of Things (IOT) are product information, electronic tag, standard expressed and uploading information. It utilises the Radio Frequency Identification (RFID) technology and wireless sensor networks (WSN). IOT applications are used in domains such as healthcare, supply chain management, defence and agriculture. Lastly the paper focuses on issues involved in IOT. Though it is a boon, IOT faces certain crucial issues like privacy and security.*

*Keywords – Internet Of Things, RFID, Electronic Tag, WSN.*

## I- INTRODUCTION

The Internet of Things is considered as the third wave of information technology right after Internet and mobile communication network, which is characterized by more comprehensive interoperability and intelligence. It was introduced by Electronic product code (EPC) Technology [1] and research work of International Telecommunication Union (ITU) . Initially there was only information online i.e. data content on the internet, then people were connected by means of e-mail, social networking but now the time has come for objects to be connected and thats what IOT achieves. It is new type of internet application and "Thing" in Internet of things refers to the product's information [1]. Hence every object be it a television or a plant, can be connected to the internet. The object's information is shared across the globe using internet and hence the objects can be accessed from a remote place [2]. The product's information is embedded in an electronic tag (RFID tag) using some standard words. The semantic meaning of these words forms ontology and hence IOT forms a part of the semantic web [3]. What is the difference between an IOT application and a normal internet application? The answer lies in two things-the way information is uploaded and the kind of information that is uploaded [1]. RFID readers are used to read the products information and then upload them onto the internet. The information uploaded has certain attributes [3] that makes it different from other applications. Moreover, the RFID objects generate a lot of dynamic sensor readings which inturn lead to frequent change of information and require more space. In contrast, WebPages are static [2], consume lesser space and they are updated weekly or monthly. The properties of IOT [1]are- it's a new type of internet application, thing's information as object , standard expression of information and non-contact uploading by a machine.

This paper discusses the growth and evolution of IOT in section II, core technologies in section III and useful applications in section IV. Lastly in section V, a brief summary of the paper with its conclusion is given.

## II- GROWTH OF IOT

Internet has been part and parcel of the social animal's life. It's a huge space of information and people. The internet first evolved as "internet of computers" [5]. It is a global platform where many services like the World Wide Web could be implemented on top of it. It was an era of information exchange. As the days passed by, people started emerging into the internet- "Internet of people" [5]. Many social websites came into picture which kept people connected all the time. This has led to internet being filled with people rather than information. On the other hand, technology has been advancing day by day and simultaneously an era of "MobiComp" (mobile computing) had begun. Mobile helped man to be always connected to the internet on the move. Nowadays 3G and 4G mobile internet connections have led to faster internet access and deliver better quality in video calls. Wireless technologies and mobile computing have become cheap and have gained more popularity [5]. Hence a new computing had emerged- Ubiquitous computing. This computing focuses on smart, intelligent space and minimal user involvement [2]. Advancement in technology led to mobile and other hand-held devices to diminish in size. Smart phones, Ipads, tablets and notebooks replaced ordinary mobiles and PCs. Hence there was a change in the device with which people access the internet. This inturn resulted in sophisticated features being configured in devices such as sensors, Global Positioning system (GPS) and actuators. In such a scenario devices were not only connected to the internet but also sense, compute and perform intelligent tasks.[5]. Later physical objects were configured with identification tags such as bar code and RFID so that they could be scanned by devices like smart phones and upload their information into the internet. This way of connecting the physical world with cyberspace with the help of a smart device led to internet being called as "Internet of Things". Hence IOT has its roots from Mobile computing, ubiquitous computing and information technology [2]. IOT connects the objects in an intelligent way. The "thing" here refers to the physical object's information read through sensors and RFID reader and uploaded into the internet. The physical object can be anything from smart phones to objects at home. The International telecommunications Union (ITU) has pointed out four dimensions of IOT : object identification (" tagging things") , sensors and wireless sensor networks(" feeling things"), embedded systems ("thinking things") and nanotechnology ("shrinking things'). Hence from the above , IOT changes the connectivity view from "**any-time , any-place**" for "**any-one**" into "**any-time , any-place**" for "**any-thing**". These things once connected to the internet provide smart services beneficial to the environment and society. They play a major role in supply chain, energy, defence, health care and other useful applications.

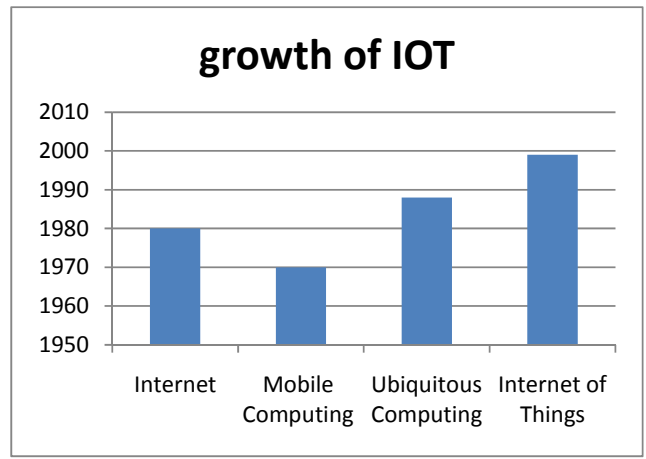

Fig 1. Growth of Internet Of Things

## III – TECHNOLOGIES

The main motive of Internet Of things is to make the things or objects in the world to be connected through internet, wireless sensor networks (WSN) and smart phones so that they can share information automatically [6] just like people sharing information. To implement this motive, there are many technologies that come into picture. Radio Frequency Identification (RFID) tags mobile phones, sensors, actuators, embedded systems and nanotechnology helps the things to communicate among themselves.

### III . A. Radio Frequency Identification

Radio Frequency identification (RFID) is a wireless technology that is used for identification of objects [6]. Due to its reduced cost and increased abilities like tracking the location, status of objects and remote reading [4], it is more preferred than the usual bar code technology. It is the root cause factor for an object to be identified so that it can be connected to the internet. RFID uses radio waves to identify things and transfer its information to the RFID reader without physical contact. The RFID system has two main components: RFID tags (transponders) and RFID Readers (transceivers) [6]. The tags have a microchip, memory to record information using Electronic Product Code (EPC) or Universal Identification (UID) and an embedded antenna. The working of an RFID application is as follows: The RFID tags are attached to the items which have to be monitored and whose information is to be shared. The readers are flashed on the tag and due to the radio signals received by the in-built antenna, the tag responses by transferring their EPC to the reader. The reader then transmits this information from EPC to the computer to be shared across the internet. In cases where smart phones are used, the sensors present in the mobile devices capture the information and are uploaded online using GPRS or Wi-Fi. Tags are of two types: active and passive. Active tags have inbuilt battery, allows reading from distance locations and transmit information frequently to the reader. On the other hand passive tags do not have a battery of their own and transmit EPC only when the transceivers come within their range [6].The above working refers to an active tag. The Passive tag responds in a different way. When the passive tag is approached by a reader, an electromagnetic signal from the reader energises the tag. Using inductive coupling [6], the energy from the signal is absorbed by the tag which converts it into electrical energy and stores in a inbuilt capacitors that it can respond to the reader with an EPC. Hence the RFID system uploads the thing's information through non-contact reading by a machine rather than humans.

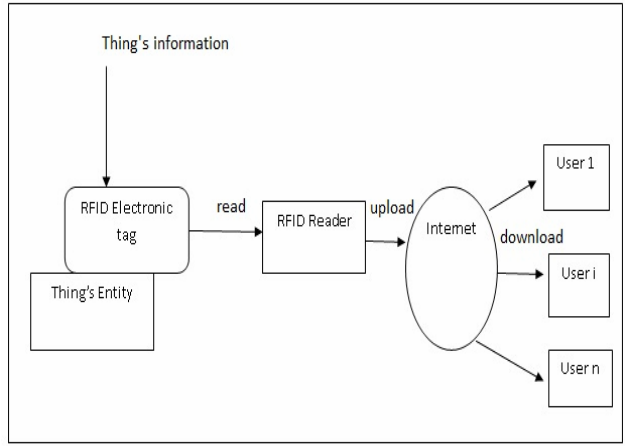

Fig .2. Radio Frequency Identification System (RFID)

**III B . Wireless Sensor Networks**

Wireless Sensor Networks (WSN) play a vital role in connecting the physical world to the information world [4].These networks monitor the changes happening in the environment; report them so that corresponding responses can be taken.
WSN help in short distance communication among the objects by building wireless networks in an ad-hoc way [7].WSN contain many independent nodes that communicate among themselves with the help of wireless radio. The nodes contain a sensor( collecting data) , microcontroller (computing data and controlling) ,memory ( storing program and data ) , radio transceiver ( for communication with other nodes)  and battery (power supply) [6]. These sensors work together to collect data and send it to the sink node. The sink node redirects the data to the destination. Hence many nodes have to coordinate together to send the signal to the sink node.

**III C. Embedded Systems and Nanotechnology**

Embedded systems are intelligent and things with embedded intelligence become smart things.  These make things perform certain actions automatically. For e.g.  A smart watching machine can wash and dry clothes automatically without human intervention. Nano-technology can imbibe intelligence in things which are called smart devices. They are able to process information, self-configure and take independent decisions [4]. These smart devices are connected with the help of LAN, GPRS, WSN, Wi-Fi, 3G, etc.

## IV. APPLICATIONS

IOT applications are used widely in many domains. Healthcare, agriculture, smart buildings (school, hospital, home) , supply chain management , Transportation and defence.

### IV.A. Agriculture

Internet of Things can be of great use in the field of agriculture. It can be helpful in monitoring growth of medicinal plants.  These plants are fitted with RFID tags and sensors. When there is a drastic or unexpected change in the growth of plant due to temperature / humidity, the sensors sense this and the RFID tags send the EPC (information) to the reader and are shared across the internet. The farmer or scientist can access this information from a remote place and take necessary actions.

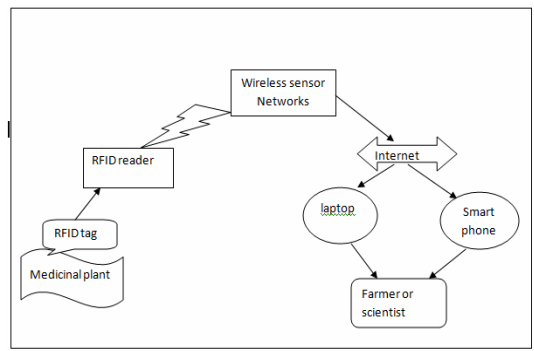

Fig .3. Agricultural system

### IV.B Smart Buildings – School

A school has many buildings in its campus like Administration block, library, Refreshment building, teaching block, etc. All these buildings have their own ventilation mechanism, AC supply and elevator systems. These facilities have to be individually managed and maintained which becomes a tedious process.  This scenario can be easily handled using Internet of Things for better management of the facilities. Each of the above blocks is fixed with RFID tag that keeps monitoring the ventilation, AC supply behaviour. The RFID system keeps sensing the change in environment and collects the data and sends it to the Information Gathering manager present in the respective block. Since the school campus will be equipped with Wi-Fi, the data from here is sent to the Central Control system. The control system on receiving the data will take necessary actions such as reducing the AC supply or stopping the elevator service. A communication mediator is required to mediate between the physical world and information world. Hence using IOT, steps are taken without human intervention.

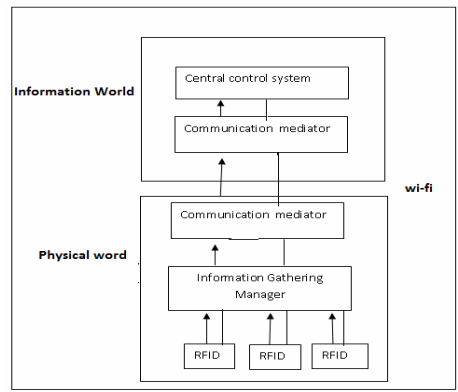

Fig. 4. School Facilities system

**IV.C Healthcare-Telemedicine**

IOT plays a crucial role in healthcare. It can be used in many ways such as tracking the number of patients in a hospital, identifying the right patient for the right medicine and monitoring a patient's health conditions from a remote place which is known as Telemedicine [8] . This includes providing treatment, diagnosis and treatment. Ambient assisted living provides technical systems for elderly people who are alone at home and need to be monitored. The patient's health status is periodically sensed using RFID and sensors. The doctor from a remote location provides medical assistance based on the information received.

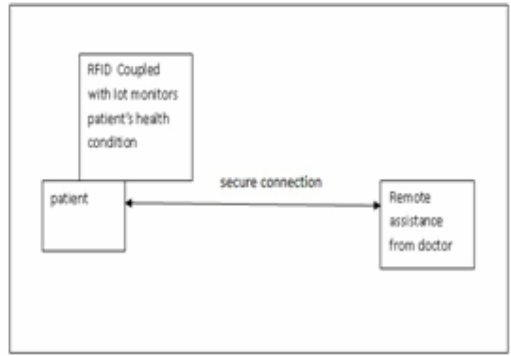

Fig.5. Telemedicine

## V. ISSUES IN IOT

Though IOT has been a boon in many ways, it also poses certain challenges. The main challenges are privacy, reliability, data confidentiality and security. A vehicle attached with RFID tag leads to lack of privacy for the passenger in the vehicle. IOT in healthcare can also lead to dangerous consequences such as the data present in the health status can be changed by an intruder , hence giving the doctor wrong information. Wireless sensors in war fields, if found by the enemies can be mishandled to generate false information. An individual's right to privacy should be protected. Strong security and sound privacy solutions will lead to better acceptance by public [4].There should be laws and policies to curb the misuse of IOT technology.  Global Standards need to be developed for the spread of this new technology.

## VI. CONCLUSION

Internet of things is a new internet application which leads to an era of smart technology where there exists **thing-thing** communication rather than human-human communication. Through IOT, each and every object in this world can be identified, connected and take decisions independently. It has taken its birth from mobile computing and ubiquitous computing. Technologies such as RFID, wireless sensor networks and embedded systems play a vital role in forming an IOT application. It is used in many applications in healthcare, agriculture, smart buildings, transportations etc. Though IOT is used in many domains, its path to success is not smooth. There are many privacy and security issues that need to be addressed. If these issues are addressed, then Internet of Things will definitely be the global mantra.

REFERENCE

[1]Yinghui Huang, Guanyu Li,"**Descriptive Models for Internet of Things"**, International Conference on Intelligent Control and Information Processing, August, 2010 - Dalian, China.

[2] Daqiang Zhang, Laurence T. Yang, Hongyu Huang, "**Searching in Internet of Things: Vision and Challenges**", Ninth IEEE International Symposium on Parallel and Distributed Processing with Applications, 2011.

[3] Yinghui Huang, Guanyu Li "**A Semantic Analysis for Internet of Things**" , International Conference on Intelligent Computation Technology and Automation , 2010.

[4] Lu Tan, Neng Wang, "**Future Internet: The Internet of Things**", 3rd International Conference on Advanced Computer Theory and Engineering (ICACTE) , 2010.

[5] Louis Coetzee, Johan Eksteen , "**The Internet of Things – Promise for the Future? An Introduction ",** IST-Africa 2011 Conference Proceedings Paul Cunningham and Miriam Cunningham (Eds) IIMC International Information Management Corporation, ISBN: 978-1-905824-24-3, 2011.

[6] Guicheng Shen, Bingwu Liu,"**The visions, technologies, applications and security issues of Internet of Things",** IEEE, 2011.

[7] Qian Zhu, Ruicong Wang, Qi Chen, Yan Liu and Weijun Qin*y,* "**IOT Gateway: Bridging Wireless Sensor Networks into Internet of Things**, IEEE/IFIP International Conference on Embedded and Ubiquitous Computing, 2010 .

[8]  A. J. Jara, M. A. Zamora and A. F. G. Skarmeta." **An ambient assisted living system for telemedicine with detection of symptoms**". Third International Work-Conference on the Interplay between Natural and Artificial Computation. Lecture Notes, pp.75-84, 2009.

[9] Ning Huansheng, and Wang Binghui, "RFID **major engineering and Internet of Things"**, Beijing: China Machine Press,  pp.13-16. (in Chinese), 2009

[10] B. Nath, F. Reynolds, and R. Want, "**RFID technology and applications**", IEEE Pervasive Computing, Vol.5, no.1, pp.22-24, 2006.

[11] R. Want, "**An introduction to RFID technology**", IEEE Pervasive Computing, Vol.5, no.1, pp. 25-33, 2006.

[12] D. Hailay and R. Roine, "**Systematic review of evidence for the benefits of telemedicine**," J Telemed Telecare , vol. 8. , pp. 1–7, 2002.

.[13] S. Misra, P. Venkata Krishna, Harshit Agarwal, Antriksh Saxena and M. S. Obaidat, "A Learning Automata Based Solution for Preventing Distributed Denial of Servicein Internet of Things", 2011 IEEE International Conferences on Internet of Things, and Cyber, Physical and Social Computing, pp. 114-122, 2011.

[14] S. Misra, P. Venkata Krishna, Harshit Agarwal, Anshima Gupta and M. S. Obaidat, "An Adaptive Learning Approach for Fault-Tolerant Routing in Internet of Things", 2012 IEEE Wireless Communications and Networking Conference: PHY and Fundamentals, pp.815-819, 2012.